\documentclass[sigconf,natbib=true]{acmart}
\usepackage{pbalance}
\usepackage{enumitem}
\usepackage{booktabs}
\usepackage{graphicx}
\usepackage{array}
\usepackage{lineno}
\usepackage{multirow}
\usepackage{hhline}
\usepackage{makecell}
\usepackage{subfigure}
\usepackage{bbm}
\usepackage[marginal]{footmisc}

\AtBeginDocument{%
  \providecommand\BibTeX{{%
    \normalfont B\kern-0.5em{\scshape i\kern-0.25em b}\kern-0.8em\TeX}}}

\setcopyright{acmcopyright}
\copyrightyear{2018}
\acmYear{2018}
\acmDOI{XXXXXXX.XXXXXXX}

\acmISBN{978-1-4503-XXXX-X/18/06}

\begin{document}

\title{
Next-Scale Generative Reranking: A Tree-based Generative Rerank Method at Meituan 
}


\author{Shuli Wang}
\authornote{Both authors contributed equally to this research.}
\authornote{Corresponding author.}
\affiliation{%
  \institution{Meituan}
   \city{Chengdu}
  \country{China}
}
\email{wangshuli03@meituan.com}

\author{Changhao Li}
\authornotemark[1]
\affiliation{%
  \institution{Meituan}
   \city{Chengdu}
  \country{China}
}
\email{lichanghao@meituan.com}

\author{Ke Fan}
\affiliation{%
\institution{Meituan}
   \city{Chengdu}
  \country{China}
  }
\email{fanke08@meituan.com}

\author{Senjie Kou}
\affiliation{%
\institution{Meituan}
   \city{Chengdu}
  \country{China}
  }
\email{kousenjie@meituan.com}

\author{Junwei Yin}
\affiliation{%
\institution{Meituan}
   \city{Chengdu}
  \country{China}
  }
\email{yinjunwei03@meituan.com}

\author{Chi Wang}
\affiliation{%
\institution{Meituan}
   \city{Chengdu}
  \country{China}
  }
\email{wangchi06@meituan.com}

\author{Yinhua Zhu}
\affiliation{%
\institution{Meituan}
   \city{Chengdu}
  \country{China}
  }
\email{zhuyinhua@meituan.com}

\author{Haitao Wang}
\affiliation{%
\institution{Meituan}
   \city{Chengdu}
  \country{China}
  }
\email{wanghaitao13@meituan.com}

\author{Xingxing Wang}
\affiliation{%
\institution{Meituan}
   \city{Beijing}
  \country{China}
  }
\email{wangxingxing04@meituan.com}

\renewcommand{\shortauthors}{Trovato and Tobin, et al.}

\begin{abstract}
In modern multi-stage recommendation systems, reranking plays a critical role by modeling contextual information. Due to inherent challenges such as the combinatorial space complexity, an increasing number of methods adopt the generative paradigm: the generator produces the optimal list during inference, while an evaluator guides the generator's optimization during the training phase.
However, these methods still face two problems. Firstly, these generators fail to produce optimal generation results due to the lack of both local and global perspectives, regardless of whether the generation strategy is autoregressive or non-autoregressive. 
Secondly, the goal inconsistency problem between the generator and the evaluator during training complicates the guidance signal and leading to suboptimal performance.
To address these issues, we propose the \textbf{N}ext-\textbf{S}cale \textbf{G}eneration \textbf{R}eranking (NSGR), a tree-based generative framework. Specifically, we introduce a next-scale generator (NSG) that progressively expands a recommendation list from user interests in a coarse-to-fine manner, balancing global and local perspectives. Furthermore, we design a multi-scale neighbor loss, which leverages a tree-based multi-scale evaluator (MSE) to provide scale-specific guidance to the NSG at each scale.
Extensive experiments on public and industrial datasets validate the effectiveness of NSGR. And NSGR has been successfully deployed on the Meituan food delivery platform.

\end{abstract}





\maketitle

\section{Introduction}
E-commerce platforms, such as Meituan and Taobao, need to provide personalized services from millions of items to their users. To improve recommendation efficiency, a typical personalized recommendation system consists of three stages: matching, ranking, and reranking. Ranking models assess items individually, with a focus on feature interactions \citeN{W&D, deepfm, xdeepfm}, user preference modeling \citeN{din, dien, sim}, among other aspects. However, these methods often overlook the critical mutual influences among contextual items and positional information.
Studies \citeN{burges2010ranknet, listnet, ai2018learning, pang2020setrank, nlgr} have shown that optimizing listwise utility during the reranking stage is a more effective strategy, as it leverages both position information and mutual influences among items in the list to improve overall performance.

\begin{figure*}[h]
\centering
\includegraphics[width=\textwidth, keepaspectratio]{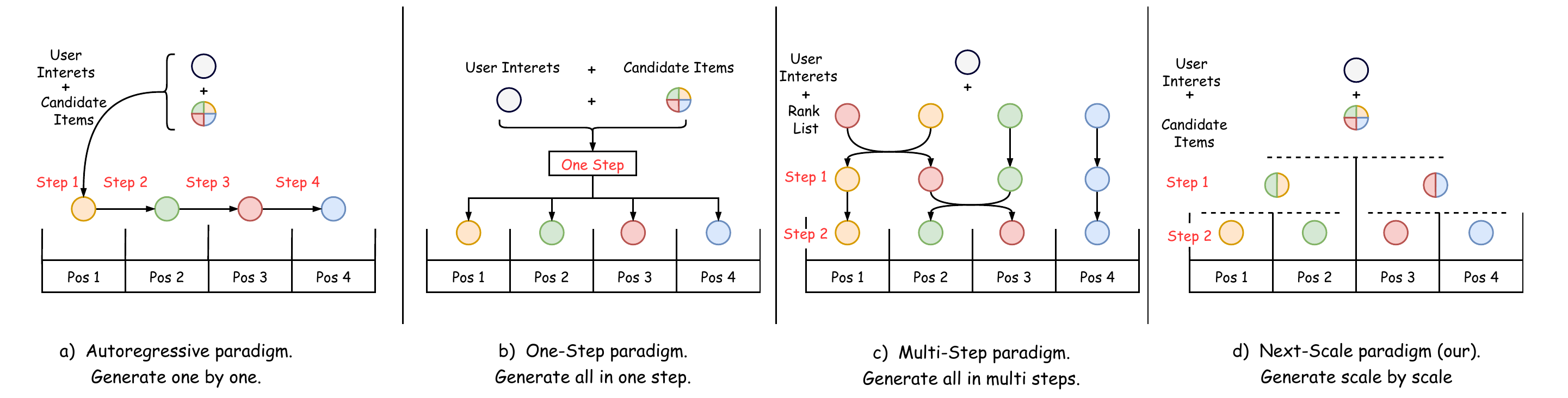}
\caption{Demo of NSGR. When generating a list of length 4, we gradually determine the position of each item from the mixed state of the 4 candidate items. 
}
\label{fig: demo}
\end{figure*}

The key challenge of reranking lies in exploring the optimal list within the vast combinatorial space \cite{gfn, nar4rec}. Early listwise methods \cite{prm, midnn, dlcm} improve scoring accuracy by modeling contextual interactions among items and then apply greedy strategies to reorder items based on these scores. However, such approaches suffer from the "evaluation-before-reranking" problem \cite{xi2021context, grn}, which prevents effective optimization in the combinatorial space.
A straightforward solution—exhaustively evaluating all possible permutations, which is globally optimal but computationally infeasible under industrial time constraints. As a result, most existing reranking frameworks adopt a two-stage architecture \cite{feng2021revisit, xi2021context, grn, nar4rec, dcdr} consisting of a generator and an evaluator. Within this paradigm, the generator plays a crucial role \cite{nar4rec}. While some methods employ heuristic generators such as SimHash \cite{chen2021end} or artificially designed diversity indicators \citeN{huang2021sliding,lin2022feature}, these often fail to leverage evaluator feedback, limiting their effectiveness. Recently, generative models \cite{grn, nar4rec, dcdr, nlgr, gref} have been adopted as generators, demonstrating superior performance compared to heuristic approaches.

Current generative reranking methods can be categorized into three types, as illustrated in Figure \ref{fig: demo}. The first category \citeN{seq2slate, dlcm, grn, prm} follows an \textbf{autoregressive paradigm}, as shown in Figure \ref{fig: demo} (a), the items are generated one by one in a left-to-right manner. This approach is not only time-consuming but also leads to suboptimal results, as the decoding process only considers previously generated items and ignores information about subsequent ones. The second category is the \textbf{one-step paradigm} \citeN{nar4rec, gref}, as shown in Figure \ref{fig: demo} (b), which generates a positional probability matrix for all items in a single step. While this approach addresses the inefficiency and lack of a global perspective seen in autoregressive methods, its overly coarse generation process increases the difficulty of modeling due to the lack of a local perspective, especially when the business is complex. The third category is the \textbf{multi-step paradigm} \citeN{dcdr, nlgr}, as shown in Figure \ref{fig: demo} (c), which starts with an initial list (e.g., the list produced by the ranking model) and iteratively swaps 1-2 items to gradually transition from the initial list to the optimal list. However, this method is prone to getting trapped in local optima due to the non-monotonic nature of the permutation space.


Beyond the limitations of the generator, how to train the generator remains a core issue. There is an inherent \textbf{goal inconsistency problem} \cite{nlgr} between the generator and the evaluator: the generator aims to find the optimal list within the combinatorial space, while the evaluator is trained to estimate list-wise scores for items. This discrepancy complicates the guidance signal from the evaluator to the generator, often resulting in the generator only fitting the exposure distribution under extreme conditions. Although NLGR \cite{nlgr} attempted to mitigate this by using relative scores of neighboring lists as guidance for the generator, it still suffers from two drawbacks: 1) it is only applicable to the multi-step generation paradigm and cannot be generalized to other paradigms; and 2) it remains susceptible to local optima due to the non-monotonic permutation space.

To address the aforementioned challenges, we propose the \textbf{N}ext-\textbf{S}cale \textbf{G}eneration \textbf{R}eranking (NSGR), a tree-based generative framework. Specifically, we introduce a next-scale generator (NSG) that starts from the user’s interests, follows a "one-generates-two, two-generate-four" pattern, and progressively refines a complete list from coarse to fine, balancing both global and local perspectives. Furthermore, we design a multi-scale neighbor loss, which leverages a tree-based multi-scale evaluator (MSE) to provide scale-specific guidance to the NSG at each scale.
As shown in Figure \ref{fig: demo} (d), NSGR operates in a coarse-to-fine manner. Taking the generation of a length-4 list as an example, the model incrementally refines the placement of items from a candidate set of four based on the user’s global interest, scaling its decisions step by step until the full list is generated.

The main contributions of our work are summarized as follows:

\begin{itemize}[leftmargin=*]
\item We propose a novel generative reranking framework featuring a next-scale progressive generation mechanism that effectively balances global and local contextual information. To the best of our knowledge, this is the first work to introduce the next-scale generation paradigm in the reranking tasks.

\item We design a multi-scale neighbor loss, which leverages a tree-based multi-scale evaluator (MSE) to guide the generator at different scales, to address the goal inconsistency problem between evaluator and generator. 

\item Extensive experiments on both public and industrial datasets demonstrate the effectiveness of NSGR, and NSGR has been successfully deployed on Meituan's food delivery platform.
\end{itemize}

\section{Related Work}
\subsection{Reranking Methods}

In recommendation systems, the reranking stage focuses on modeling contextual information and identifying the optimal list from the permutation space. Existing reranking approaches can be systematically divided into two main categories \cite{gfn}: generator-based methods \cite{prm, mir, grn} and evaluator-based methods \cite{feng2021revisit, pier}.

Generator-based methods directly produce a final list by capturing mutual influences among items. Also referred to as one-stage methods \cite{pier, nar4rec}, these approaches originally relied on behavioral logs for training. For example, Seq2Slate \cite{seq2slate} employs a pointer network, while MIRNN \cite{zhuang2018globally} uses a GRU to sequentially determine item order. Methods such as PRM \cite{prm} and DLCM \cite{ai2018learning} take an initial ranked list as input, apply RNN or self-attention to model contextual signals, and output a score for each item. However, such methods often suffer from the "evaluation-before-reranking" problem \cite{xi2021context}, leading to suboptimal performance. Similarly, EXTR \cite{extr} estimates the predicted CTR (pCTR) for each candidate item in every candidate position, but remains essentially a point-wise model with limited ability to capture fine-grained context. MIR \cite{mir} attempts to capture set-to-list interactions through a permutation-equivariant module.

Evaluator-based methods, on the other hand, aim to assess every possible permutation using a well-designed context-aware model. Due to strict latency requirements in online systems, most evaluator-based methods adopt a two-stage architecture that includes a filtering stage \cite{pier, feng2021revisit} or generating stage \cite{nar4rec} before evaluation to reduce the candidate set size. For instance, PRS \cite{feng2021revisit} uses beam search to generate a small set of candidate permutations, which are then scored by a permutation-wise ranking model. PIER \cite{pier} applies SimHash \cite{charikar2002similarity, chen2021end, manku2007detecting} to select top-K candidates from the full permutation set. In industrial recommender systems, the concept of evaluator-based methods is broader, as any generator-based approach can be used in the first stage within a multi-channel retrieval framework \cite{nar4rec}. A key limitation, however, is the inconsistency between the two stages, which restricts overall effectiveness. YOLOR \cite{yolor} addresses this issue in smaller permutation spaces by using tree-based enumeration with contextual caching to exhaustively evaluate permutations, though its applicability remains limited by the size of the permutation space.

\subsection{Generative Reranking Solutions}

Since it is difficult to achieve the optimization of the permutation space in supervised training, some generator-based methods using evaluators \citeN{listcvae, grn} have become popular in recent years. For example, ListCVAE \cite{listcvae} utilizes conditional variational autoencoders (CVAE) to capture the positional biases of the elements and the interdependencies within the list distribution. GRN \cite{grn} proposes an evaluator-generator framework to replace the greedy strategy. Still, it cannot avoid the evaluation-before-reranking problem \cite{xi2021context} because it takes the rank list as input to the generator. DCDR \cite{dcdr} introduces diffusion models into the reranking stage and presents a discrete conditional diffusion reranking framework. NAR4Rec \citeN{nar4rec} uses a non-autoregressive generative model to speed up sequence generation. NLGR \cite{nlgr} proposes the use of neighbor loss to guide the generator's training. Even though these generative methods leverage the capabilities of the evaluator, the consistency issue between the generator and the evaluator remains unresolved.



\begin{figure*}[htbp]
\begin{center}
\includegraphics[width=\textwidth]{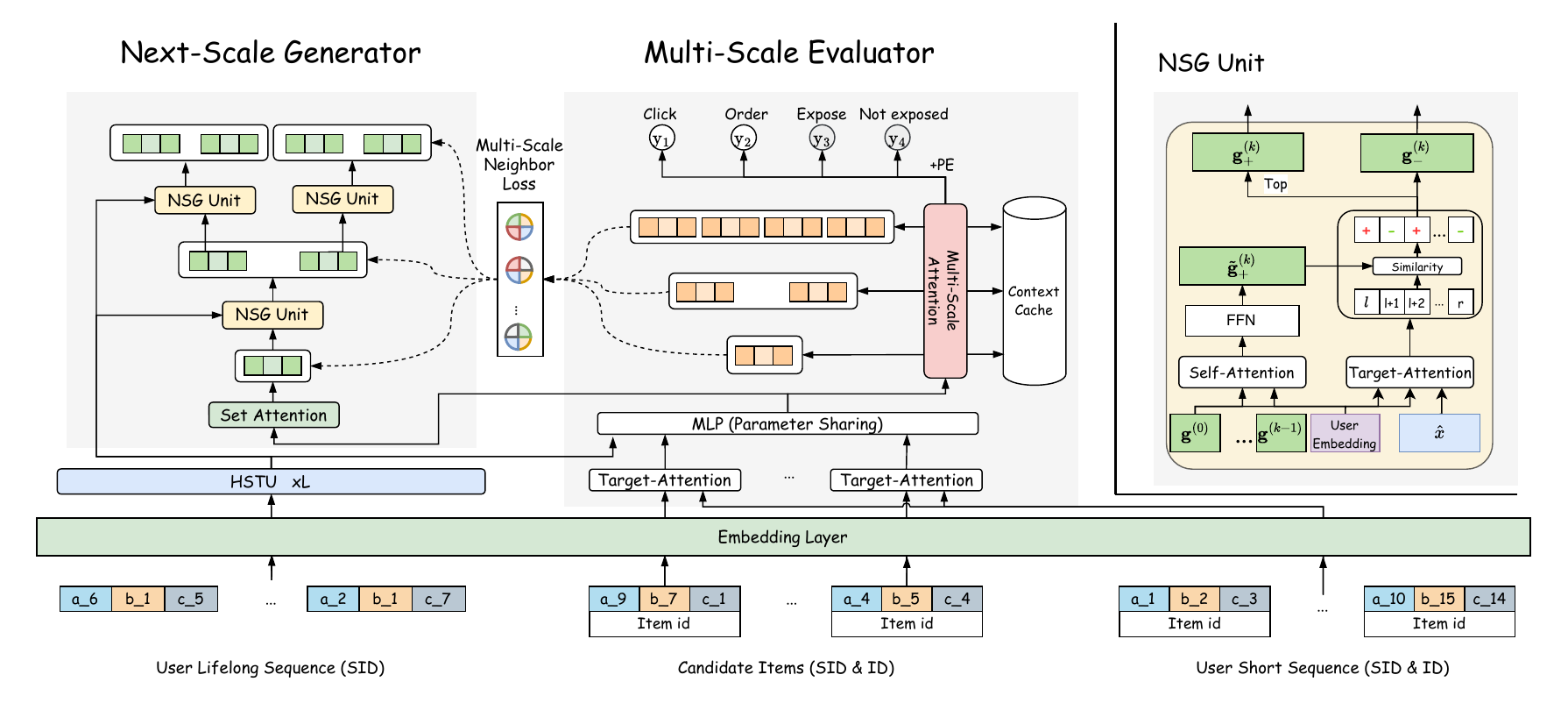}
\end{center}
\caption{The overall architecture of NSGR.}
\label{fig:nsgr}
\end{figure*}

\section{Problem Definition}

Let $ U = \{ u_1, u_2, \ldots, u_{|U|} \} $  represent a set of $|U|$ users which consist of some profile features (e.g. user ID, gender, age) and $ X = \{ x_1, x_2, \ldots, x_{|X|} \} $  represent a set of $|X|$ items. For each user $u$, given a candidate set with $n$ items $S = \{ x_1, x_2, \ldots, x_n \} $, 
the reranking model needs to recommend an ordered list with $m$ ($m \leq n$) items $L = \{ x_1, x_2, \ldots, x_m \} $ from $\mathcal{O}(A_n^m)$ candidate space to the user.

\noindent \textbf{Evaluator Objective:} The evaluator aims to accurately estimate the listwise utility  $\mathcal{R}(u, L)$ for each ordered list $L$: 
\begin{equation}\label{eq:evaluator}
E^*=\mathop{\arg\min}\limits_{E}  \mathcal{L}\left(E(u,L), \mathcal{R}(u,L)\right),
\end{equation}
where $\mathcal{L}$ denotes the loss funtion.

\noindent \textbf{Generator Objective:} The generator seeks to produce the optimal list $L^*$ with the highest utility value over the permutation space:
\begin{equation}\label{eq:generator}
G^*=\mathop{\arg\max}\limits_{G}E^*(G(u, S)),
\end{equation}
\begin{equation}\label{eq:generator2}
L^*=G^*(u, S).
\end{equation}

\section{Proposed Method}

This section details the architecture of the proposed NSGR framework. As illustrated in Figure \ref{fig:nsgr}, NSGR consists of two core components: (1) the \textbf{Next-Scale Generator (NSG)} module, which employs a coarse-to-fine strategy to produce optimal lists (Section \ref{section: Generator});  (2) and the \textbf{Multi-Scale Evaluator (MSE)} module, which captures multi-scale contextual information to accurately assess list value (Section \ref{section:mse}).
The training procedure for the NSG module is subsequently elaborated in Section \ref{sec:training}. Each component is comprehensively described in its respective subsection.

\subsection{Multi-Scale Evaluator}\label{section:mse}

The Multi-Scale Evaluator (MSE) module models multi-scale list contextual relationships based on users' semantic interests. First, we extract the user's global interest $\mathbf{e}_u$ by collecting their lifelong behavior sequence: $\mathcal{H}_u = \{x_1, x_2, ..., x_H\}$, where $H$ represents the length of the user's lifelong behavior sequence. Inspired by generative recommendation, we employ a variant of the transformer—the Hierarchical Sequential Transduction Unit (HSTU) \cite{hstu}—to extract the user's semantic interest $\mathbf{e}_u$: 
\begin{equation}\label{eq:hstu}
\mathbf{e}_u=\mathrm{AvgPool}\left(\mathrm{HSTU}(\mathcal{H}_u)\right),
\end{equation}
where $\mathrm{AvgPool}$ is the average pooling operation. Specifically, each user behavior $x_i$ is mapped to a three-level semantic ID \cite{tiger, das, rpg} to enhance generalization capability. Furthermore, to reduce computational costs during inference, $\mathbf{e}_u$ is precomputed offline using the Next Token Prediction (NTP) training approach.


Then, we enhance the semantic representation of candidate items using the user's global interest. Let $\mathbf{X} = \{\mathbf{x}_1, \mathbf{x}_2, ..., \mathbf{x}_n\} \in \mathbb{R}^{n \times D}$ represent embeddings of candidate items, $\mathcal{H}_u^{short} = \{x_1, x_2, ..., x_s\}$ represent the user’s recent $s$ behaviors, and $\mathbf{e}^o$ denote the embeddings of other features, where $D$ is the dimension of the embedding layer. We use an MLP as a simple feature-crossing unit to extract the semantic embedding for each candidate item:
\begin{equation}\label{eq:xs}
\mathbf{x}^s_i=\mathrm{MLP}\left(\mathbf{x}_i||\mathrm{TA}(\mathbf{x}_i,\mathcal{H}_u^{short}) ||\mathbf{e}^u||\mathbf{e}^o\right), \forall i \in [n],
\end{equation}
where $\mathbf{X}_i \in \mathbb{R}^{D}$ is the $i$-th candidate item in $\mathbf{X}$, || represents concatenate operate, and $\mathrm{TA}$ represents target attention operate. 

So we can obtain the list semantic vector $\mathbf{L}^s$ of each candidate list:  
\begin{equation}\label{eq:ls}
\mathbf{L}^s=\{\mathbf{x}^s_1, \mathbf{x}^s_2, ..., \mathbf{x}^s_m\}.
\end{equation}

Subsequently, we extract multi-scale context information of $\mathbf{L}^s$. For any position $t \in [m]$, its multi-scale context is calculated by multi-scale Self-Attention (SA) \cite{vaswani2017attention}:
\begin{equation} \label{eq: mse}
\begin{aligned}
&\mathbf{e}^{(1)}_t = \mathbf{e}_{1,m} = \mathrm{SA}(\mathbf{x}^s_1 || \mathbf{x}^s_{2} || ... || \mathbf{x}^s_m),
\\
&\mathbf{e}^{(2)}_t = \mathbf{e}_{1,m/2} = \mathrm{SA}(\mathbf{x}^s_1 || ... ||\mathbf{x}^s_{t} || ... || \mathbf{x}^s_{m/2}),
\\
&...
\\
&\mathbf{e}^{(K)}_t = \mathbf{e}_{t,t+1} = \mathrm{SA}(\mathbf{x}^s_t || \mathbf{x}^s_{t+1}),
\end{aligned}
\end{equation}
where $K =\log_{2}{m}$. Note that the SA layer in the above formula does not have position encoding, which can increase reusability and reduce calculations. Then we collect multi-scale contextual relationships associated with $x_t$, denoted as $\mathbf{x}^c_t=[\mathbf{e}^{(1)}_t;\mathbf{e}^{(2)}_t;\ldots;\mathbf{e}^{(\log_{2}{m})}_t] \in \mathbb{R}^{\log_{2}{m} \cdot D}$.

Finally, the position-aware CTR prediction integrates three information sources:
\begin{equation}\label{eq:8}
\hat{y}_t = \sigma\Big(\mathrm{MLP}\big(
\underbrace{\mathbf{x}^s_t}_{\text{semantics}} || \underbrace{\mathbf{x}^c_t}_{\text{context}} || \underbrace{\mathbf{e}^p_t}_{\text{position}}
\big)\Big),
\end{equation}
where $\mathbf{e}^p_t \in \mathbb{R}^D$ is the positional embedding and $\sigma$ is the Sigmoid Function. The list-wise value aggregates all positional predictions:
\begin{equation}\label{eq:9}
\hat{y}_L = \sum_{t=1}^m \hat{y}_t.
\end{equation}
This formulation allows convenient adjustment of scoring metrics according to business needs, such as Impression Rate (IMPR), Conversion Rate (CVR) and Gross Merchandise Volume (GMV).

\subsection{Next-Scale Generator}

The Next-Scale Generator (NSG) module generates the optimal
list within the combinatorial space via a hierarchical coarse-to-fine
process. At each step $k$, the NSG operates on the candidate subset
$S^{(k)}_{l:r} = \{\mathbf{x}^s_l, \mathbf{x}^s_{l+1}, \ldots, \mathbf{x}^s_{r}\}$ corresponding
to the position interval $[l:r)$, which could be obtained from Eq. \ref{eq:xs}, and splits it into two child subsets
by computing priority-aware interactions among items before
alignment with the pre-vector.

\paragraph{Item Priority.}
For each item $\mathbf{x}^s_i \in S^{(k)}_{l:r}$, we compute a priority score:
\begin{equation}
p_i^{(k)} = \mathrm{MLP}_p(\mathbf{x}^s_i) \in \mathbb{R},
\end{equation}
where a higher $p_i$ indicates greater individual relevance.

\paragraph{Pairwise Relationship Classification.}
Each ordered pair $(i, j)$ within $S^{(k)}_{l:r}$ is classified into
\emph{competitive suppression}, \emph{complementary enhancement},
or \emph{neutral coexistence}:
\begin{equation}
\mathbf{r}_{ij}^{(k)} = \bigl[r_{ij}^{\mathrm{sup}},\;
r_{ij}^{\mathrm{enh}},\;
r_{ij}^{\mathrm{neu}}\bigr]
= \mathrm{softmax}\!\left(
\mathrm{MLP}_{\mathrm{rel}}\bigl(
[\mathbf{x}^s_i;\; \mathbf{x}^s_j;\; \mathbf{x}^s_i - \mathbf{x}^s_j;\; \mathbf{x}^s_i \odot \mathbf{x}^s_j]
\bigr)
\right)
\end{equation}
The three components are mutually exclusive and sum to one.

\paragraph{Asymmetric Influence Weight.}
The directed influence of item $i$ on item $j$ is:
\begin{equation}
w_{ij} =
- r_{ij}^{\text{sup}} \cdot \text{ReLU}(p_i - p_j)
+ r_{ij}^{\text{enh}} \cdot \frac{p_i + p_j}{2}.
\end{equation}
Suppression is asymmetric (higher-priority suppresses lower), while enhancement is symmetric. Neutral pairs contribute $w_{ij} \approx 0$.

\paragraph{Set-Conditioned Item Refinement.}
We condition each item's refinement on both pairwise interactions
and the subset-level context $g^{(k-1)}$:
\begin{equation}
\hat{x}^s_j = \mathbf{x}^s_j + W_\Delta\,\mathrm{MLP}_{\mathrm{set}}\!\left(
\Bigl[\, \mathbf{x}^s_j \;;\;
\textstyle\sum_{i \neq j} w_{ij}^{(k)} \mathbf{x}^s_i \;;\;
g^{(k-1)} \,\Bigr]
\right),  \forall x^s_j \in S^{(k)}_{l:r}.
\end{equation}

\paragraph{Item-to-Tree Attention Scoring and Binary Split.}

We first generate the anchor vector $\tilde{g}^{(k)}$ by aggregating the refined representations through self-attention, capturing a \emph{global set-level} view of the current candidate subset:
\begin{equation}
\tilde{g}^{(k)} = \mathrm{FFN}\!\left(
\mathrm{SA}\bigl(e_u \,\|\, g^{(0)} \,\|\, \cdots \,\|\,
g^{(k-1)} \bigr)
\right).
\end{equation}
To further capture \emph{item-specific} relevance, each refined item $\hat{x}^s_j$ queries the user interest and historical tree nodes via target attention, producing a personalized context vector:
\begin{equation}
a^{(k)}_j = \mathrm{TA}\!\left(
\hat{x}^s_j,\;
\bigl[e_u \,\|\, g^{(0)} \,\|\, \cdots \,\|\, g^{(k-1)}\bigr]
\right).
\end{equation}
The relevance score is computed by fusing the anchor vector and
the item-specific context:
\begin{equation}
\mathrm{Sim}^{(k)}_j = \mathrm{MLP}_{\mathrm{score}}\!\left(
\bigl[\hat{x}^s_j \;;\; a^{(k)}_j \;;\; \tilde{g}^{(k)}\bigr]
\right).
\end{equation}
The candidate subset is split by ranking $\mathrm{Sim}^{(k)}_j$:
\begin{align}
\mathrm{Flag}^{(k)}_+ &= \mathbf{1}\!\left[
\mathrm{rank}(j) < \tfrac{r-l}{2}
\right] \\
g^{(k)}_+ &= \mathrm{AvgPool}\!\left(
\hat{x}^s_j \cdot \mathrm{Flag}^{(k)}_+
\right) \\
g^{(k)}_- &= \mathrm{AvgPool}\!\left(
\hat{x}^s_j \cdot (1-\mathrm{Flag}^{(k)}_+)
\right)
\end{align}
This process recurses for $K = \log_2 m$ steps, yielding
$g^{(K)} \in \mathbb{R}^{m \times D}$.

\begin{figure}[h]
\centering
  \includegraphics[width=\linewidth, height=1\textheight, keepaspectratio]{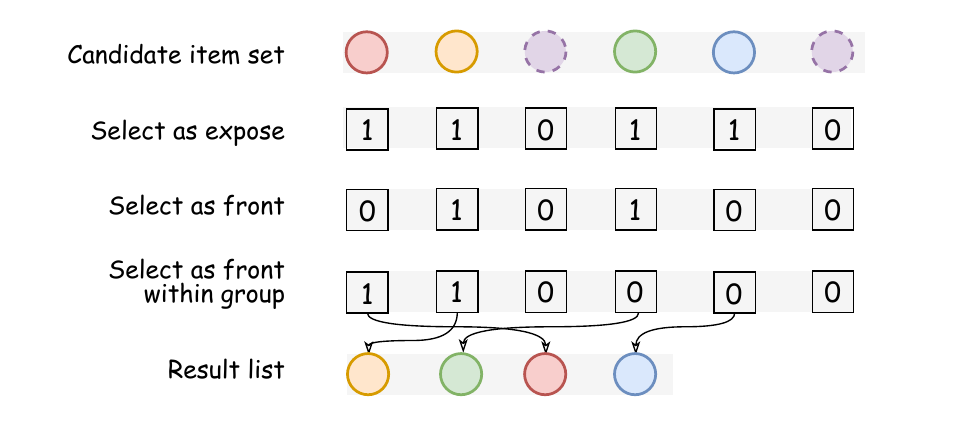}
\caption{Next-scale generation process demo.}
\label{fig: sort}
\end{figure}

\paragraph{Final Ranking.}
The final ranked list can be obtained by aggregating the \texttt{Flag}$^{(k)}_+$ masks accumulated at each step of the tree traversal, as illustrated in Figure \ref{fig: sort}.

\subsection{Training with Multi-Scale Neighbor Loss}\label{sec:training}


As mentioned before, the goal inconsistency between the evaluator and the generator complicates the transfer of guidance. We will introduce our solution in detail below.




\subsubsection{\textbf{Training of the MSE}}
The MSE is tuned to fit the list-wise value of items. We train the MSE using real data collected from online logs, including exposure, click, conversion, and other performance indicators. The loss of MSE is calculated as follows:

\begin{equation}
\mathcal{L}_E = -\sum_{t=1}^m \left[ y_t \log(\hat{y}_t) + (1 - y_t) \log(1 - \hat{y}_t) \right],
\end{equation}
where subscript $t$ is the index of response items, $y_t$ represents the real label, $\widehat{y}_{t}$ represents the predicted value, $m$ is the length of the pageview list, which includes unexposed items.

\subsubsection{\textbf{Training of NSG}}
The NSG is aligned to generate the optimal list within the permutation space. We leverage the MSE as the reward model, to guide the alignment of NSG. However, directly training the generator is challenging due to the sparsity of the permutation space. Inspired by NLGR \cite{nlgr} and GRPO \cite{grpo}, we propose to use multi-scale neighbor loss (MSNL) to construct relative rewards for the generator training.

\begin{figure}[h]
\centering
  \includegraphics[width=\linewidth, height=1\textheight, keepaspectratio]{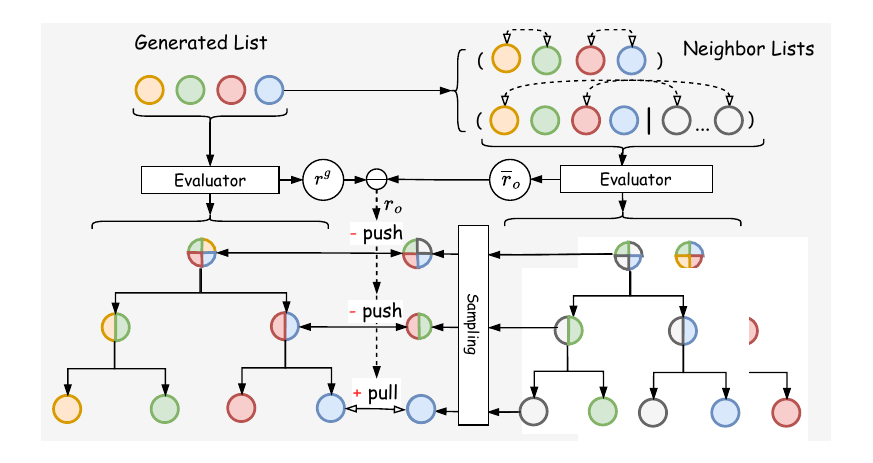}
\caption{Multi-scale neighbor loss demo.}
\label{fig:loss}
\end{figure}

First, we collect neighbor lists at multiple scales. As illustrated in Figure \ref{fig:loss}, a list of length 4 contains 7 nodes in its tree structure. Therefore, to ensure adequate training of NSG, the construction of neighbor lists needs to consider each node across multiple scales.
For convenient representation, we define the list generated by NSG as $L^g$, and the neighbor lists as $\overline{L}=\{\overline{L}_1,\overline{L}_2,..., \overline{L}_{O}\}$, where $O$ denotes the number of neighbor lists. These neighbor lists include two types: one involves swapping positions of items within $L^g$, and the other involves swapping positions between items in $L^g$ and other candidate items.

Then, we use MSE to evaluate the list-wise values of $L^g$ and the neighbor list $\overline{L}$, represented as $r^g$ and $\overline{r}=\{\overline{r}_1,\overline{r}_2,..., \overline{r}_{O}\}$, respectively. We use the relative values between $L^g$ and the neighbor lists to guide the training of NSG, expressed as:
\begin{equation}
r_o=\overline{r}_o-r^g, \forall o \in [O].
\end{equation}

Note that since NSG and MSE have similar tree structures, we can directly use MSE’s $\mathbf{e}^{(k)}$ (in Eq. \ref{eq: mse}) to guide NSG’s $\mathbf{g}^{(k)}$ (in Eq. \ref{eq: align}), expressed as:
\begin{equation}\label{loss: nsg}
\mathcal{L}_G = -\sum_{k=1}^K \sum_{o=1}^O \mathrm{log} \frac{\mathbbm{1}_{r_o > 1} r_o \cdot\mathrm{exp}\left({\mathbf{g}_o^{(k)}}^\top\mathbf{e}_o^{(k)}/\tau\right)}{\sum_{o=1}^O\mathbbm{1}_{r_o < 1} \mathrm{exp}\left({\mathbf{g}_o^{(k)}}^\top\mathbf{e}_o^{(k)}/\tau\right)},
\end{equation}
where $\mathbbm{1}_{(\cdot)}$ is 1 when $(\cdot)$ is true, else 0. In addition, the multi-scale vectors of neighbor lists can also be saved to the Context Cache to avoid repeated calculations.

The parameters of MSE are frozen when training NSG, and the HSTU module is pretrained with the Next-Token Prediction task.

\section{Experiments}
To validate the superior performance of NSGR, we conducted extensive offline experiments on the Meituan dataset and verified the superiority of NSGR in online A/B tests. In this section, we first introduce the experimental setup, including the dataset and baseline. Then we present the results and analysis of various reranking methods in both offline and online A/B tests.

\subsection{Experimental Setup}
\subsubsection{Dataset}
In order to verify the effectiveness of NSGR, we conduct sufficient experiments on both public dataset and industrial dataset. For public dataset, we choose Taobao Ad dataset. For industrial dataset, we use real-world data collected from Meituan food delivery platform. Table \ref{tab:my_table} gives a brief introduction to the datasets.

\begin{table}[hbt!]
\caption{Statistics of datasets.}
\label{tab:my_table}
\begin{tabular}{cccc}
\hline
\textbf{Dataset} & \textbf{\#Users} & \textbf{\#Items} & \textbf{\#Records} \\ \hline
Taobao Ad               & 1,141,729        & 99,815           & 26,557,961           \\
Meituan          & 5,685,119      & 17,264,613       & 242,549,848        \\ \hline
\end{tabular}
\end{table}

\begin{itemize}[leftmargin=*]
\item  Taobao Ad \footnote{https://tianchi.aliyun.com/dataset/56}. It is a public dataset collected from the display advertising system of Taobao. This dataset contains more than 26 million interaction records of 1.14 million users within 8 days. Each sample comprises five features: user ID, timestamp, behavior type, item brand ID, and category ID. It includes four behavior types: browse, cart, like, and buy, and each behavior is timestamped. We use the first 7 days as training samples (20170506-20170512), and the 8th day as test samples (20170513).
\item Meituan. It is an industrial dataset collected from the Meituan food delivery platform during August 2025, which contains 242 million interaction records of 5.6 million users within 15 days. The dataset includes 239 features, three labels: expose, click and conversion, and collects all items on the same page as one record. We use the data of the first 14 days as the training set, and the data of the last 1 day as the test set.
\end{itemize}

Note that all samples are list-level, that is, each sample contains all items in a request list. We filter out samples whose labels are all 0 or all 1.

\subsubsection{Baseline}
The following six state-of-the-art ranking and reranking methods are chosen for comparative experiments and divided into four groups. We select PRM and GRN as the autoregressive paradigm baselines (Group I), NAR4Rec as the one-step paradigm baseline (Group II), DCDR and NLGR as the multi-step paradigm baselines (Group III). YOLOR as evaluator-based baseline method (Group IV). A brief introduction of these methods is as follows:

\begin{itemize}[leftmargin=*]
\item $\textbf{PRM}$\cite{prm}, autoregressive paradigm, adjusts an initial list by applying the self-attention mechanism to capture the mutual influence between items.
\item $\textbf{GRN}$\cite{grn}, autoregressive paradigm, is a generative reranking model which consists of the
evaluator for predicting interaction probabilities and the generator for generating reranking results.
\item $\textbf{NAR4Rec}$\cite{nar4rec}, one-step paradigm, presents a non-autoRegressive generative model for reranking Recommendation.
\item $\textbf{DCDR}$\cite{dcdr}, multi-step paradigm, presents a discrete conditional diffusion reranking framework.
\item $\textbf{NLGR}$\cite{nlgr}, multi-step paradigm, proposes a utilizing neighbor lists model for Generative Reranking.
\item $\textbf{YOLOR}$\cite{yolor} utilizes a tree-based extraction and context cache to perform full calculations in the permutation space of $A_8^8$.

\end{itemize}

\subsubsection{Evaluation Metrics. }
We employ multiple metrics to comprehensively evaluate NSGR in offline experiments. These include \textbf{AUC} (Area Under the ROC Curve) and \textbf{GAUC} (Group AUC), where the latter is adapted from standard AUC to better suit reranking scenarios by computing the average AUC within each list. High AUC and GAUC values indicate the model's effectiveness in ranking positive samples above negative ones, both globally and at the intra-list level, reflecting its discriminative capacity across different contexts. For online experiments, we use \textbf{CTR} (Click-Through Rate), \textbf{CVR} (Conversion Rate), \textbf{GMV} (Gross Merchandise Volume) and \textbf{time-cost (ms)} as key performance indicators.

To assess the consistency between the generator and evaluator stages, we adopt \textbf{HR (Hit Ratio)} \cite{alsini2020hit}. For a given input, HR equals 1 only if the generator produces a permutation set that contains the optimal list. Detailed results and analysis of the consistency evaluation are provided in Section \ref{section:consistency}.

It is important to note that AUC/GAUC and HR capture complementary aspects of reranking performance. AUC/GAUC reflects the evaluator’s ability to accurately assess the quality of an ordered list, whereas HR measures the generator’s effectiveness in maintaining consistency between the two stages. A deficiency in either metric can impair overall recommendation performance.


\subsubsection{Implementation Details}
We implement all the deep learning baselines and NSGR with TensorFlow 1.15.0 using NVIDIA A100-80GB GPU. For all comparison models and our NSGR model, we adopt Adam as the optimizer with the learning rate fixed to 0.001 and initialize the model parameters with a normal distribution by setting the mean and standard deviation to 0 and 0.01, respectively. The batch size is 512, the embedding size is 8. The hidden layer sizes of MLP in Eq. \ref{eq:8} are (1024, 256, 128). For the Taobao Ad dataset, the length of the ranking list and reranking list are both 4, thus the length of the full permutation is 24. For Metuan dataset, we rerank 20 items from the initial ranking list which contains 20 candidate items, thus the length of the full permutation is $A_{20}^{20}=2.43 \times 10^{18}$.
All experiments are repeated 5 times, and the averaged results are reported.

\subsection{Performance of the Evaluator}\label{exp_result}

Here we show the results of our proposed method NSGR. All results are averaged from 5 experiments. As can be seen in Table \ref{tab:2} and Table \ref{tab:3}, NSGR outperforms baselines, including several recent reranking methods.
We have the following observations from the experimental results: 
1. All re-ranking listwise models (e.g. PRM, GRN) make great improvements over point-wise models (e.g. DNN, DeepFM) by modeling the mutual influence among contextual items, which verifies the impact of context on user click behavior.
2. Compared with generator-based methods(e.g. PRM), evaluator-generator methods also improve the CTR prediction because they evaluate more candidate lists. 
3. Our proposed NSGR brings 0.0045/0.0153 absolute AUC and 0.0047/0.0160 absolute GAUC on Taobao/Meituan dataset, gains over the state-of-the-art independent baseline, which is a significant improvement in the industrial recommendation system.

\begin{table}[h]
\caption{Comparison between NSGR and baseline methods on the Taobao Ad dataset.}
\label{tab:2}
\begin{tabular}{l|ccc}
\toprule
Model & AUC & GAUC & Loss   \\
\midrule
PRM         & 0.6052 & 0.8163 & 0.1842   \\
GRN         & 0.6101 & 0.8209 & 0.1820   \\
NAR4Rec     & 0.6306 & 0.8288 & 0.1786   \\
DCDR        & 0.6217 & 0.8288 & 0.1792   \\
NLGR        & 0.6344 & 0.8311 & 0.1752   \\
YOLOR       & \underline{0.6351} & \underline{0.8323} & \underline{0.1743}\\
NSGR (Our)        & \textbf{0.6396} & \textbf{0.8389} & \textbf{0.1713}\\
\bottomrule
\end{tabular}
\end{table}

\begin{table}[h]
\caption{Comparison between NSGR and baseline methods on Meituan.}
\label{tab:3}
\begin{tabular}{l|ccc}
\toprule
Model & AUC & GAUC & Loss   \\
\midrule
PRM     & 0.8595 & 0.8573 & 0.1008 \\
GRN     & 0.8643 & 0.8598 & 0.1001 \\
NAR4Rec & 0.8711 & 0.8636 & 0.0957 \\
DCDR    & 0.8695 & 0.8616 & 0.0977   \\
NLGR    & 0.8732 & 0.8644 & 0.0946   \\
YOLOR   & \underline{0.8749} & \underline{0.8669} & \underline{0.0932}\\
NSGR (Our)   & \textbf{0.8902} & \textbf{0.8829} & \textbf{0.0842}\\
\bottomrule
\end{tabular}
\end{table}

\subsection{Performance of the Generator}\label{section:consistency}
We compare the HR (Hit Ratio) of PRM, GRN, NAR4Rec, YOLOR and NSGR on Meituan datasets ($A_{20}^{20}$). Due to the enormous permutation space of the Meituan dataset, evaluating all possible permutations is infeasible, which also poses a challenge to the consistency metrics. We adopt an estimation approach for HR@1\% and HR@10\%, which is done by comparing the generator's target list against 1,000 randomly sampled candidate lists.
As shown in Figure \ref{tab:4}, we have the following observations: 1. NSGR achieves the highest HR@1\% (0.86) and HR@10\% (0.99), significantly outperforming all baseline methods. 2. YOLOR performs second-best, with HR@1\% and HR@10\% reaching 0.82 and 0.93, respectively. Among the earlier methods, NAR4Rec shows better consistency than GRN and PRM. 3. The results demonstrate that NSGR exhibits superior consistency in generating high-quality lists, validating the effectiveness of the proposed next-scale generation paradigm.

\begin{table}[h]
\caption{HR performance comparison on Meituan.}
\label{tab:4}
\begin{tabular}{lcccccc}
\toprule
Model & PRM & GRN & NAR4Rec & NLGR & YOLOR & NSGR   \\
\midrule
HR@1\%     & 0.510 & 0.632 & 0.658 & 0.784 & \underline{0.822} & \textbf{0.861} \\
HR@10\%    & 0.691 & 0.844 & 0.897 & 0.916 & \underline{0.943} & \textbf{0.987} \\
\bottomrule
\end{tabular}
\end{table}

\begin{figure}[h]
\centering
\includegraphics[width=\linewidth, height=1\textheight, keepaspectratio]{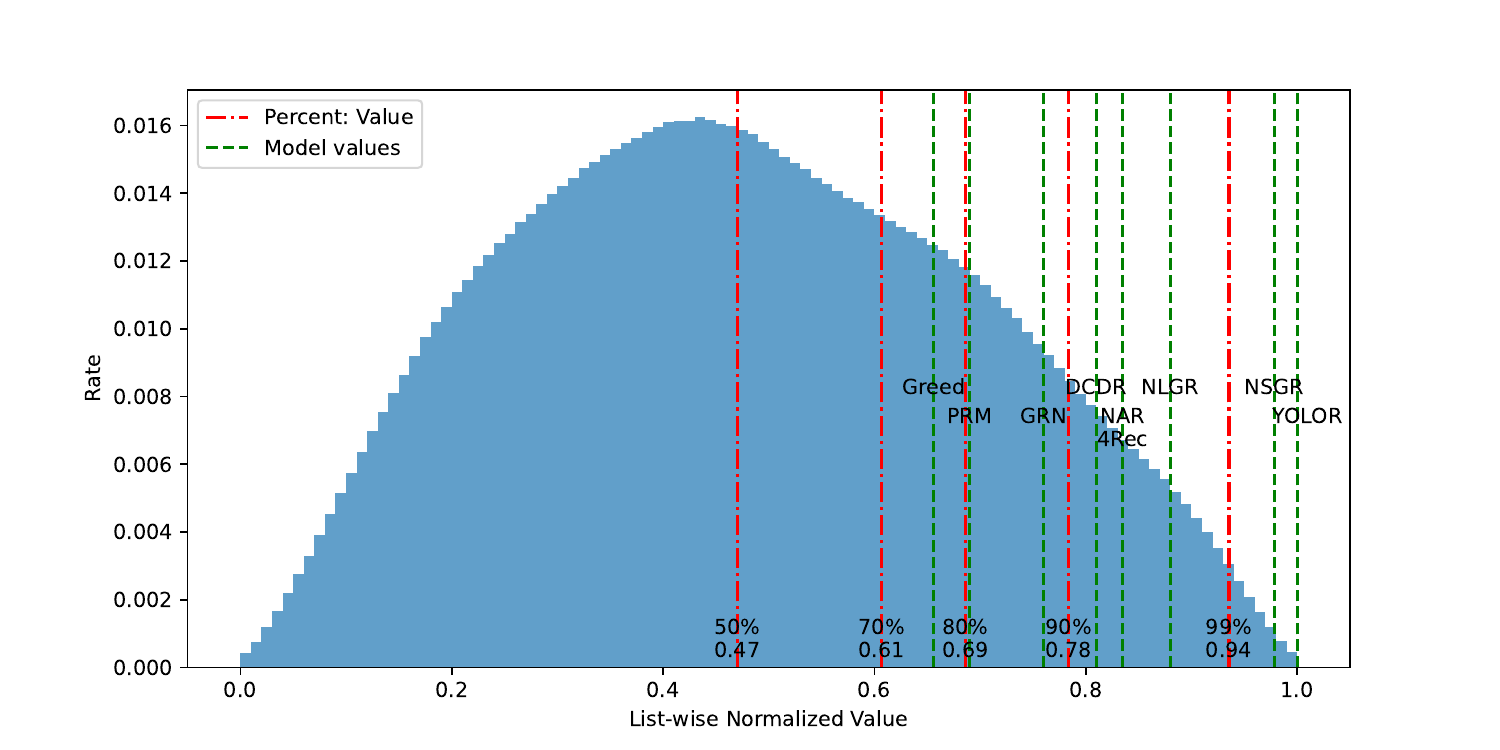}
\caption{Performance comparison within the $A_8^8$ permutation space distribution.}
\label{fig:hist}
\end{figure}

Furthermore, to demonstrate the consistency advantage of NSGR more vividly, we collect 2,000 users' request logs and build a dedicated test set within the $A_8^8=40,320$ permutation space. By evaluating every candidate list in $A_8^8$ space, we assess the performance of those reranking methods as authentically as possible. The blue histogram in Figure \ref{fig:hist} displays the normalized value distribution for all $A_8^8$ lists, averaged over 2,000 users. The red line denotes the [50\%, 70\%, 80\%, 90\%, 99\%] percentile values, and the green line the comparative model's value. As shown in Figure \ref{fig:hist}, we have the following observations: 1. The value distribution in the permutation space approximates a normal distribution. 2. The greedy ordering (Greed line) based on ranking scores yields a value of only 0.66, proving large gains from reranking. 3. NSGR achieves a superior value of 0.978, demonstrating a considerable performance lead (YOLOR achieves the optimal value of 1 because we reduce the permutation space from $A_{20}^{20}$ to $A_8^8$, enabling it to exhaustively evaluate the entire permutation space).

\begin{figure*}[htbp]
\begin{center}
\includegraphics[width=0.9\textwidth]{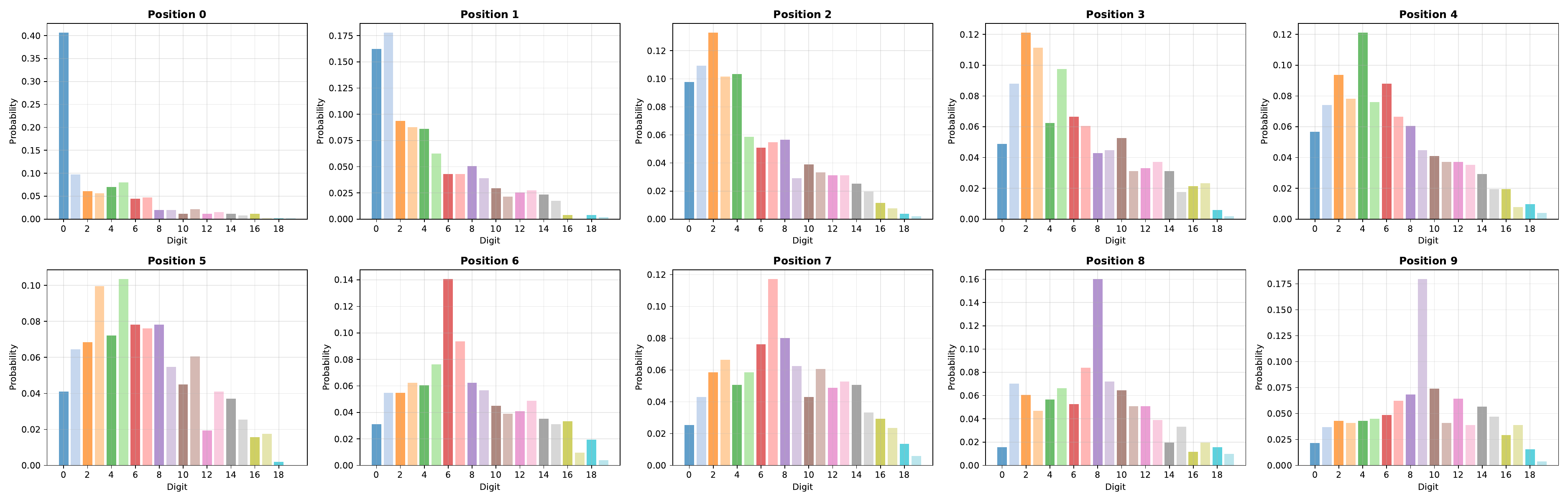}
\end{center}
\caption{Positional distribution of Top-10 items following NSGR.}
\label{fig:position}
\end{figure*}

As shown in Table \ref{tab:5}, we evaluate the proximity of the list generated by NSGR to the optimal list of $A_8^8$. While the exact match rate (Same) is 0.689, the accuracy increases significantly to 0.909, 0.933, and 0.968 when allowing for total deviation counts of 2, 3, and 4 items, respectively. This confirms that NSGR produces lists that are very close to the optimal permutation.

\begin{table}[h]
\caption{Comparison between the list generated by NSGR and the optimal list.}
\label{tab:5}
\begin{tabular}{l|cccc}
\toprule
Model & Same & $\mathrm{Diff}_2$ & $\mathrm{Diff}_3$ & $\mathrm{Diff}_4$ \\
\midrule
NSGR  & 0.689 & 0.909 & 0.933 &  0.968 \\
\bottomrule
\end{tabular}
\end{table}

\subsection{Result Position Analysis}

We proceed with a further analysis of the ranking position changes produced by the NSGR. As shown in Figure \ref{fig:position}, each subplot shows the probability that an item in a given final position originated from each of the 20 initial positions. The positional distribution analysis reveals three fundamental characteristics: 1. Each ranking position predominantly retains its originally top-ranked item, demonstrating persistent positional inertia. 2. Allocation probabilities exhibit a consistent decay pattern as positional displacement increases, adhering to a distance-decay relationship. 3. The distribution morphology transitions from sharply peaked dominance in position-$0$ toward flattening distributions in posterior ranks. This progression indicates the presence of diminishing marginal returns: earlier positions in the list yield substantially greater utility gains than later ones.

\subsection{Ablation Study}
To assess the effectiveness of each component in NSGR, we conducted a series of ablation studies using the Taobao Ad and Meituan datasets. Specifically, we build several variants of the NSGR:
\begin{itemize}[leftmargin=*]
\item w/o SID. A variant of NSGR that replaces the SID and HSTU with item id and DIN \cite{din}.
\item w/o MSE unit. A variant of NSGR without the MSE unit (MSEU), which is replaced by a single global self-attention.
\item w/o NSG unit. A variant of NSGR without the NSG unit (NSGU), which is replaced by a single softmax.
\item w/o MSNL. A variant of NSGR that replaces the $\mathcal{L}_G$ in Eq. \ref{loss: nsg} with $r^g$.
\end{itemize}

\begin{table}[h]
\caption{The contributions of different components of NSGR.}
\label{tab:6}
\begin{tabular}{l|ccc}
\toprule
Model &  AUC & GAUC & HR@1\% \\
\midrule
w/o SID           & 0.8761 & 0.8692 & 0.834 \\
w/o MSEU          & 0.8835 & 0.8742 & 0.846 \\
w/o NSGU          & 0.8902 & 0.8829 & 0.796 \\
w/o MSNL          & 0.8902 & 0.8829 & 0.772\\
NSGR              & \textbf{0.8902} & \textbf{0.8829} & \textbf{0.861} \\
\bottomrule
\end{tabular}
\end{table}
Table \ref{tab:6} shows the results of the ablation study, and we can draw the following conclusions: 
1. The absence of the SID leads to a significant performance drop in both the evaluator and the generator.
2. The MSE unit is crucial for the evaluator, indicating that the multi-scale context possesses stronger context extraction capabilities compared to a single self-attention mechanism.
3. The NSG unit is vital for the generator's performance, but does not affect the evaluator's performance.
4. The HR of NLGR w/o NSNL drops the most, demonstrating that MSNL is essential for the generator.

\subsection{Hyperparameter Analysis}
We analyze the sensitivity of two hyperparameters: $\tau$ and $\beta$, corresponding to the generation process and training process of NLGR. Among them, $\tau$ is the temperature coefficient in Eq. \ref{loss: nsg}, and $\beta$ is the sampling ratio at each position when constructing the neighbor lists $\overline{L}$. By default, $\beta=1$ means that each position is sampled $1$ time, that is $O=m$. The result is shown in Table \ref{tab:tab4}, showing the same trend on the public dataset and industrial dataset and we have the following findings:
\begin{enumerate}[leftmargin=*]
\item [i)] As the hyperparameter $\tau$ increases within a certain range, the HR of NSGR maintains a stable level, first increasing and then decreasing.

\item [ii)] We tested several values for $\beta$. When $\beta<1$, we randomly select $\beta m$ positions in $m$ positions to construct rewards. When $\beta>1$, we construct $\beta$ neighbor lists at each position. Increasing $\beta$ within a certain range can quickly improve the HR performance. As $\beta$ continues to increase, HR remains stable but increases offline training time. The results show that counterfactual rewards considering all positions are important.
\end{enumerate}

\begin{table}[h]
\caption{Parameter sensitivity analysis on Meituan.}
\label{tab:tab4}
\begin{tabular}{llccccc}
\hline
& $\tau$=0.01 & $\tau$=0.1 & $\tau$=0.5 & $\tau$=1.0 & $\tau$=2.0  \\
\hline
HR@1\%     &  0.842   &  \textbf{0.861}   &  0.858   &  0.851    &   0.849   \\
HR@10\%    &  0.977   &  \textbf{0.987}   &  0.983   &  0.980    &   0.979   \\
\hline
   & $\beta$=0.1 & $\beta$=0.5 & $\beta$=1 & $\beta$=2 & $\beta$=5 \\
\hline
HR@1\%      &  0.823   &   0.842  &  0.859   &  \textbf{0.861}    &   0.860   \\
HR@10\%    &  0.951   &  0.975   &  0.986   &  \textbf{0.987}    &   0.987   \\
\hline
\end{tabular}
\end{table}

\subsection{Performance on Online System} \label{subsection:5.6}
\begin{figure}[hbtp]
\centering
\includegraphics[width=\linewidth, height=1\textheight, keepaspectratio]{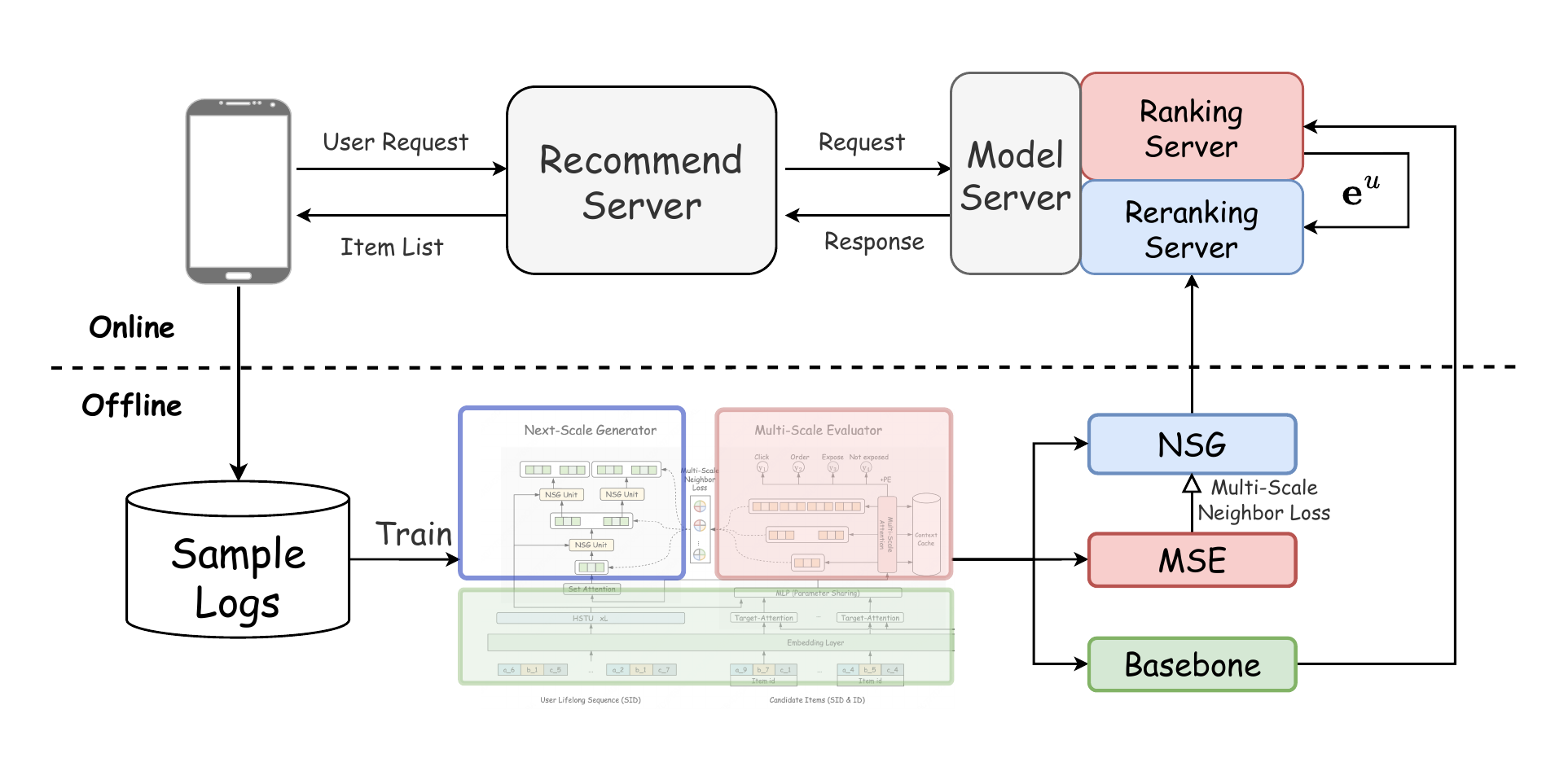}
\caption{Architecture of the online deployment with NSGR.}
\label{fig:nsgr_online}
\end{figure}

To evaluate the online performance of NSGR, we deployed NSGR on the Meituan CPS business as shown in Figure \ref{fig:nsgr_online}. It is worth noting that because the point-wise part of NSGR is exactly the ranking model, NSGR can directly reuse the $\mathbf{e}^u$ and $\mathbf{x}^s$ output by the ranking model to reduce calculations and increase link consistency.

We also conducted a rigorous A/B test for eight weeks, from August 2025 to October 2025. Specifically, we assigned NSGR with 30\% traffic, while the remaining 70\% traffic was assigned to baseline -- YOLOR(8) with $A_8^8$ permutation space.
Table \ref{tab:online} shows the online performance of NSGR, where NSGR(8) represents NSGR with $A_8^8$ permutation space, and NSGR(20) represents NSGR with $A_{20}^{20}$ permutation space. 
Compared to the baseline model (YOLOR), although NSGR (8) shows a slightly negative effect, when the number of candidate items increases from 8 to 20, NSGR (20) achieves a 2.89\% improvement in CTR and a 3.15\% increase in GMV, representing a highly significant growth for the business.
Now, NSGR is deployed on the Meituan food delivery platform and serves millions of users.

\begin{table}[h]
\caption{Online A/B test result.}
\label{tab:online}
\begin{tabular}{c|cccc}
\toprule
\textbf{Method}        & \textbf{CTR} & \textbf{CVR} & \textbf{GMV} & \textbf{Cost(ms)}  \\ 
\midrule
\textbf{NSGR ( 8 )}        &  -0.42\%     & -0.18\%      & -1.02\%      & -2.1  \\
\textbf{NSGR (20)}         &  +2.89\%     & +0.58\%      & +3.15\%      & -1.4  \\
\bottomrule
\end{tabular}
\end{table}

\section{Conclusion}

In this paper, we propose the Next-Scale Generation Reranking (NSGR), a tree-based generative framework. Specifically, NSGR utilizes the Next-Scale Generator (NSG), a tree-based generator that progressively expands a recommendation list from user interests in a coarse-to-fine manner—ensuring a balance between global relevance and local coherence. We further design a multi-scale neighbor loss, which leverages a tree-based multi-scale evaluator (MSE) to provide scale-specific guidance to the NSG at each scale.
Extensive experiments on public and industrial datasets demonstrate the effectiveness of NSGR, which has been successfully deployed on Meituan's food delivery platform.


\bibliographystyle{ACM-Reference-Format}
\bibliography{main}

\end{document}